\def\be{\begin{eqnarray}}
\def\ee{\end{eqnarray}}
\def\ps{p\hspace{-0.075in}/}
\documentstyle[12pt]{article}
\begin{document}

\begin{flushright}
USITP-92-11\\
October 1992
\end{flushright}
\bigskip
\Large
\begin{center}
\bf{Discretization of the Superparticle Path Integral}\\

\vspace{2.5cm}

\normalsize
by\\
\bigskip

J.Grundberg\\
{\it Department of Theoretical Physics\\
The Royal Institute of Technology\\
S-100 44 Stockholm\\
SWEDEN\\
\bigskip
and\\}
\bigskip
U.Lindstr\"om and H.Nordstr\"om\\
{\it ITP\\
University of Stockholm\\
Vanadisv\"agen 9\\
S-113 46 Stockholm\\
SWEDEN}\\
\end{center}
\vspace{3.0cm}
\normalsize
{\bf Abstract:} Requiring that the path integral has the
global symmetries of the classical action and obeys the natural composition
property of path integrals, and also that the discretized action has the
correct
naive continuum limit, we find a viable discretization of the (D=3,N=2)
superparticle action.

\eject
In this letter we discuss the path integral quantization of a N=2
superparticle
in three dimensions.

There have been numerous attempts to quantize both the
massive and the massless superparticle \cite{EVAN1}-\cite{MIKO}. Both the
massless and massive models are invariant under a certain fermionic symmetry;
the
Siegel symmetry \cite{SIEG1}. For the massive case
quantization has been carried out both using BV-BRST methods \cite{GREE} and
using covariant methods \cite{EVAN1}. In the massless case a covariant
separation of the models first and second class constraints is not possible in
general \cite{BENG}. Attempts to circumvent this problem have been
made using BV-BRST methods \cite{LIND1} and using harmonic
superspace methods \cite{NISSI}. Also non-covariant quantization has been
described \cite{EVAN2}. A constructive path integral quantization has, to our
knowledge, only been attempted in \cite{MIKO}, however.
Our construction starts
from a set of (natural) requirements on the path integral; it should have the
global symmetries of the classical action and it should obey the usual
composition property of a path integral. In addition, we require the
discretized
action to have the correct naive continuum limit.  We find a discretization
that complies with these demands and construct the propagator. This propagator
differs from the usual field theory propagator as well as from the propagator
derived in \cite{MIKO}. Nevertheless, as we show in a forthcoming publication
\cite{GRUN}, calculations of physical entities yield the same result as in
field
theory.

Our analysis highlights that the path integral is a formal object
which has to be given content by some evaluation prescription. This is
particularly the case for fermionic variables. We also want to draw
attention to the fact that the relation between the first quantized theory
and the field theory is not always as simple as in the ordinary scalar field
case. This is worth having in mind when trying to find a string field theory.

The point of departure is the $N=2, D=3$ superparticle phase space action:
\be
S=\int\limits_0^1 {d\tau \left\{ {p_\mu \left[ {\dot x^\mu +{\textstyle{i
\over 2}}\dot {\bar \theta} \Gamma ^\mu \theta -{\textstyle{i \over 2}}\bar
\theta \Gamma ^\mu \dot \theta } \right]-{\textstyle{i \over 2}}m(\dot
{\bar \theta } \theta -\bar \theta \dot \theta )-e (p^2+m^2)}
\right\}}. \label{action} \ee
Here $\theta _\alpha $ is a complex anticommuting $3D$ spinor, $(\alpha
=1,2)$, and the Dirac matrices satisfy
\be
\left\{ \Gamma ^\mu,\Gamma ^\nu \right\}=-2\eta ^{\mu \nu}
\ee
with metric signature $(-,+,+)$. The action (\ref{action}) is invariant under
the (global) $N=2$ supersymmetry,
\be
\delta \theta _\alpha &=&\varepsilon _\alpha\cr
\delta \bar \theta ^\alpha &=&\bar \varepsilon ^\alpha\cr
\delta x^\mu &=& {\textstyle{i \over 2}}(\bar \varepsilon \Gamma ^\mu \theta
-\bar \theta \Gamma^\mu \varepsilon)\cr
\delta p_\mu &=& 0\cr
\delta e &=& 0  \label{susy}
\ee
reparametrizations,
\be
\delta \theta _\alpha &=&\xi \dot \theta _\alpha \cr
\delta \bar \theta ^\alpha &=&\xi \dot {\bar \theta} ^\alpha \cr
\delta x^\mu &=& \xi \dot x ^\mu \cr
\delta p_\mu &=& \xi \dot p _\mu \cr
\delta e &=& \dot {(\xi e)} \label{rep}
\ee
and Siegel symmetry
\be
\delta \theta  &=&(\ps + m)\kappa\cr
\delta \bar \theta &=&\bar \kappa(\ps +m)\cr
\delta x^\mu &=& {\textstyle{i \over 2}}(\bar \theta \Gamma ^\mu \delta \theta
- \delta \bar \theta \Gamma^\mu \theta)\cr
\delta p_\mu &=& 0\cr
\delta e &=& -i(\dot{\bar \theta }\kappa-\bar \kappa \dot \theta).
\label{Sieg}
\ee
To find the propagator we should evaluate
\be
K=\int{{\it DeDpDxD\theta D\bar \theta}exp \left\{{iS}\right\}}.\label{K}
\ee
As usual, we fix the reparametrizations by choosing
\be
\dot e = 0 \quad \Rightarrow \quad e=T.
\label{repg}
\ee
The formal expression (\ref{K}) is then replaced by the slightly less formal
\be
K=\int \limits _0^\infty {dTG}
\label{intG}
\ee
where
\be
G(x _f,\theta _f,\bar \theta _f;x _i,\theta _i,\bar \theta _i;T)
=\int{{\it DpDxD\theta D\bar \theta}exp\left\{ iS \right\}}\label{G}
\ee
is a function of final and initial superspace positions and
\be
S=\int\limits_0^T {d\tau \left\{ {p_\mu \left[ {\dot x^\mu +{\textstyle{i
\over 2}}\dot{\bar \theta }\Gamma ^\mu \theta -{\textstyle{i \over 2}}\bar
\theta
\Gamma ^\mu \dot \theta } \right]-{\textstyle{i \over 2}}m(\dot{\bar \theta}
\theta -\bar \theta \dot \theta )-(p^2+m^2)} \right\}}.
\label{action2} \ee
The main body of this paper is concerned with giving a meaning to (\ref{G}),
(which we will henceforth call the propagator).

One way of defining a path integral is to discretize, i.e., to put the theory
on a (time-) lattice. There are infinitely many ways to do this. For bosonic
variables the usual choice is to let, (see, e.g., \cite{SCHU}),
\be
x(t) \to {{x(t_{i+1})+x(t_i)} \over 2} \equiv
{{x_{i+1}+x_i} \over 2}\cr
\dot x(t) \to  {{x(t_{i+1})-x(t_i)} \over {t_{i+1}-t_i}} \equiv
{{x_{i+1}-x_i} \over \varepsilon}.
\label{xd}
\ee
If we try this prescription for our fermionic $\theta$'s, we find an
ambiguity;
the result depends on whether the number $N$ of time steps is even or odd (see
also \cite{MIKO}). In \cite{MIKO} the rule (\ref{xd}) with odd $N$ was
choosen for fermions. We feel that this is unacceptable. A basic intuitive
property of a path integral is that one should be able to calculate the
amplitude of going from $A$ to $C$ by first calculating the amplitude from $A$
to $B$, then that from $B$ to $C$ and finally summing over intermediate
postitions $B$ \cite{FEYN}. The path integral as defined in \cite{MIKO} does
not
fulfil this. This  property is also an important ingredient in perturbative
calculations. It leads to the  structure "$ (propagator)\times
(vertex)\times(propagator)\times...$", a structure familiar from field theory.

Guided by the above considerations we require that $G$ in (\ref{G}) satisfies
\be
G(3;1;T_1+T_2)=\int{d^3x_2d^2\theta _2d^2 \bar \theta _2G(3;2;T_2)G(2;1;T_1)}
\label{cmp}
\ee
(where the argument has been abbreviated, e.g., $x_3,\theta _3,\bar \theta _3
\to 3$ et.c.). The condition (\ref{cmp}) both implies a restriction on the
possible discretizations and helps us determine the path integral measure. We
suggest the following form for the infinitesimal propagator:
\be
G(f;i;\varepsilon )=\int{{{d^3p} \over {(2\pi )^3}}{1 \over
{p^2+m^2}}exp\left\{iS\right\}} \label{infpro}
\ee
\eject
where
\be
S&=&p_\mu \left[{x ^\mu _f-x ^\mu _i
-{\textstyle{i \over 2}}\bar \theta _f \Gamma ^\mu(\theta _f-\theta _i)
+{\textstyle{i \over 2}}(\bar \theta _f -\bar \theta _i)\Gamma ^\mu\theta _i
}\right]\cr
&&-(p^2+m^2)\varepsilon-{\textstyle{i \over
2}}m\left[{(\bar \theta _f -\bar \theta _i)\theta _i-\bar \theta _f(\theta
_f-\theta _i) }\right]. \label{discs}
\ee
(The unusual factor ${\textstyle{1 \over {p^2+m^2}}}$ is dictated by the
composition rule (\ref{cmp}) and will be further commented on below.) The
discretized path integral becomes,
\be &&G(f;i;T)=\mathop{lim}\limits_{N\to
\infty}\int {\prod_{k=1}^{N-1} {\left( {{{d^3p_kd^3x_kd^2\theta _kd^2\bar
\theta
_k} \over {(2\pi )^3(p_k^2+m^2)}}} \right){{d^3p_N} \over {(2\pi
)^3(p_N^2+m^2)}}
}}\times\cr
&&\exp \left\{ {i\sum\limits_{k=1}^N {\left( {-(p_k^2+m^2)\varepsilon
-{\textstyle{i \over 2}}m\left[ {(\bar \theta _k-\bar \theta _{k-1})\theta
_{k-1}-\bar \theta _k(\theta _k-\theta _{k-1})} \right]} \right.}}
\right.\cr
&&+\left.{\left.{p_\mu ^k\left[ {x_k^\mu -x_{k-1}^\mu -{\textstyle{i
\over 2}}\bar \theta _k\Gamma ^\mu (\theta _k-\theta _{k-1})+{\textstyle{i
\over
2}}(\bar \theta _k-\bar \theta _{k-1})\Gamma ^\mu \theta _{k-1}}
\right]}\right)}\right\}
\label{discpi}
\ee
but since (\ref{infpro}) satisfies the composition property (\ref{cmp}) we
merely have to replace $\varepsilon \to T$ in (\ref{infpro}) to evaluate
(\ref{discpi}) and thus obtain the finite form.

The discretization (\ref{discs})
has several virtues: It has the correct naive continuum limit, it is
translationally invariant and it is space-time supersymmetric, (c.f.
(\ref{susy})). The asymmetry between $\theta$ and $\bar \theta$ should be
noted. We remind the reader that a similar asymmetry exists in the path
integral
approach to fermions using coherent states \cite{OHNU}.
A further property of (\ref{discpi}) is that it can be written as, (dropping
the
$m$-dependence)
\be
&&G(f;i;T)\cr
&&={\textstyle{1 \over
2}}D_f^\alpha D_f^\beta \bar D_\beta ^i\bar D_\alpha
^i\left[{\delta^4(\theta _f-\theta _i)\int{{d^3p \over {(2\pi)^3}}{1 \over
{p^2}}exp\left\{{-iTp^2+ip_\mu (x_f^\mu -x_i^\mu
)}\right\}}}\right]\cr
&&.\label{Ddelta}
\ee
where
\be
D^\alpha \equiv {\partial \over {\partial \theta _\alpha}}-{\textstyle{i \over
2}}\left({\bar \theta \Gamma ^\mu}\right)^\alpha {\partial \over {\partial
x^\mu }}\cr
\bar D_\alpha \equiv {\partial \over {\partial \bar \theta
^\alpha}}-{\textstyle{i \over 2}}\left({\Gamma ^\mu\theta}\right)_\alpha
{\partial \over {\partial x^\mu }}.\label{Covder}
\ee
This means that the propagator is (anti-)chiral with respect to the (final)
initial point in superspace. $\bar D^i$ in (\ref{Ddelta}) could just as well
be
replaced by $\bar D_f$. Using the $p^{-2}$ under the integral sign, we thus
have the antichiral projection operator ${\textstyle{{D^2\bar D^2} \over
{p^2}}}$
acting. In fact, a projection operator is precisely what is needed to obtain
the composition property (\ref{cmp}). Clearly an interchange $\bar\theta
\leftrightarrow \theta$ leads to another viable discretization. This
corresponds to replacing the antichiral projection operator by the chiral one.
As a third and final independent projection operator we may use the linear
operator $\propto {\textstyle D\bar D^2D}$. The corresponding discretized
action
contains extra $\theta$-terms that vanish in the naive continuum limit.

Compared to the propagator derived in \cite{SIEG1},\cite{MIKO}, we note the
following discrepancies: First, all alternatives described above differ in the
derivative structure from the propagator in \cite{SIEG1},\cite{MIKO}. Further,
our
propagator has an additional factor $(p^2+m^2)^{-1}$, which is surprising also
compared to the superfield propagators. However, the physical meaning of the
propagator is unclear at this stage. To test the validity of our expression we
need to calculate some physical process. The calculations will be presented
elsewhere \cite{GRUN}, here we just stress some general features.  We
consider couplings of the superparticle to background gauge fields.The Siegel
invariant coupling is:
\be S=\int\limits_0^1 {d\tau \left\{{(p_\mu -A_\mu
)\pi ^\mu -e(p^2+(A+m)^2) }\right.}\cr \left.{-i\dot{\bar \theta}^\alpha \bar
A_\alpha -iA^\alpha \dot \theta _\alpha -{\textstyle{i \over
2}}m(\dot{\bar \theta}\theta - \bar \theta \dot{\theta})}\right\}\label{scoup}
\ee
where
\be
\pi ^\mu \equiv \dot x^\mu -{\textstyle{i \over 2}}\left({\bar \theta \Gamma
^\mu \dot \theta -\dot{\bar \theta}\Gamma ^\mu \theta}\right)\label{pi}
\ee
and, for completeness, we have included a mass. The gauge multiplet is
\be
A_\mu &\equiv &{\textstyle{1 \over2}}(\Gamma _\mu )_\alpha ^\beta [D^\alpha
,\bar
D_\beta]V\cr
\bar A_\alpha &\equiv &\bar D_\alpha V\cr
A^\alpha &\equiv &D^\alpha V\cr
A &\equiv &iD^\alpha \bar D_\alpha V \label{gaugem}
\ee
where $V$ is a superfield prepotential. We derived (\ref{scoup}) from the
formalism developed in \cite{GAUN}. It should be compared to the $10D$
coupling presented in \cite {ROCE}. Note that, unlike in the superfield
theory,
the coupling involves the dimensionful gauge potentials rather than the
dimensionless prepotential. This discrepancy between the superparticle and
superfield coupling to a gauge field matches the discrepancy in the
propagators. The factor $p^{-2}$ in the propagator (\ref{Ddelta}) ensures that
it
has the right dimension for the perturbation expansion to be dimensionally
consistent. This follows from the composition rule (\ref{cmp}) and that the
action is dimensionless. In fact, as will be shown in \cite{GRUN}, a suitable
discretization of (\ref{scoup}), (with $m=0$), allows us to recover the usual
superfield theory results for a physical process.

These last considerations have all been for the massless case. Including a
mass
will correspond to  $N=2$ supersymmetry with a central charge. The expression
(\ref{Ddelta}) changes in that the covariant derivatives include the central
charge and in that $p^2 \to p^2 +m^2$.

As mentioned in the introduction, the massive Siegel invariant superparticle
has
been quantized (in various dimensions) using BRST-methods \cite{GREE}, as well
as
canonical ones \cite{EVAN1}, (without calculating the propagator however). No
difficulties of principle were encountered. For the massless case, the
canonical
procedure is faced with the difficulty that a covariant separation of first
and
second class constraints is impossible \cite{BENG}. This has led to
quantization using Batalin-Vilkovisky type Lagrangian BRST methods
\cite{LIND1}.
These constructions involve an infinite tower of ghosts. Our construction
would
seem to circumvent both these difficulties. We have ignored the Siegel
invariance, although the (anti)chirality with respect to the endpoints may be
viewed as a remnant of this symmetry. The issue of first and second class
constraints never arises, the construction involves no ghosts, and the limit
$m
\to 0$ seems unproblematic. Our treatment of the path integral may seem to be
particular to $D=3, N=2$, but it is clear that this construction of the
propagator works just as well for the $D=4, N=1$ massless superparticle. What
about other cases?

The assumptions in this
letter are, (explicitly), the composition property (\ref{cmp}) and,
(implicitly), that the exponent in the infinitesimal propagator should look
"reasonable" as a discretization of the action. In a conventional treatment of
the path integral with gauge fixing and ghosts, a modified version of
(\ref{cmp})
involving also ghost coordinates should be satisfied, since this is
essentially
the completeness property of intermediate states. Truncating to
$(x^\mu,\theta,
\bar \theta)$ is thus an assumption that the ghost coordinates decouple, which
is reasonable. The second assumption is more questionable. Gauge fixing would
certainly modify the action and with the gauge condition $\dot{\theta}=0$
\cite{SIEG2}, the $\theta$-dependent part of the propagator would be $\delta
^N(\theta _f-\theta _i)$, (which certainly satisfies the composition
property).
The role of the ghosts is to remove unphysical states from this, i.e., to
project onto an irreducible representation of supersymmetry. This is precisely
what the projection operator in our construction accomplishes. (C.f. the
situation in string theory with ghosts versus the Brink-Olive projection
operator \cite{GSW}). It thus seems that the existence of a projection
operator
that projects onto an irreducible representation of supersymmetry is the basic
requirement in our construction.

\vspace{2.5cm}

\begin{flushleft}
{\bf Acknowledgement:} We thank H.Hansson, A.Karlhede, M.Ro\v cek and W.Siegel
for numerous discussions on the subject of this article.
\end{flushleft}
\eject

 \end{document}